\documentclass[aps,pra,amsmath,amssymb,twocolumn,superscriptaddress]{revtex4}
\pdfoutput=1 
\usepackage[usenames]{xcolor}
\usepackage[pdftex]{graphicx}
\usepackage{amsmath, amstext, amssymb, amsfonts, amsxtra}
\usepackage{xspace}
\usepackage{blindtext}
\usepackage{braket}
\usepackage{bbold}
\usepackage{here}
\usepackage[colorlinks=true,linkcolor=black,citecolor=blue,urlcolor=black]{hyperref}
\usepackage{siunitx}

\begin{document}
\title{Quantum nanofriction in trapped ion chains with a topological defect }
\author{L. Timm}
\email[]{lars.timm@itp.uni-hannover.de}
\affiliation{Institut f\"ur Theoretische Physik, Leibniz Universit\"at Hannover, Appelstr. 2, 30167 Hannover, Germany}
\author{L. A. Rüffert}
\affiliation{Physikalisch-Technische Bundesanstalt, Bundesallee 100, 38116 Braunschweig, Germany}
\author{H. Weimer}
\affiliation{Institut f\"ur Theoretische Physik, Leibniz Universit\"at Hannover, Appelstr. 2, 30167 Hannover, Germany}
\author{L. Santos}
\affiliation{Institut f\"ur Theoretische Physik, Leibniz Universit\"at Hannover, Appelstr. 2, 30167 Hannover, Germany}
\author{T. E. Mehlst\"aubler}
\affiliation{Physikalisch-Technische Bundesanstalt, Bundesallee 100, 38116 Braunschweig, Germany}
\affiliation{Institut f\"ur Quantenoptik, Welfengarten 1, 30167 Hannover}

\date{\today}
\begin{abstract}
Trapped ion systems constitute a well controllable scenario for the study and emulation of nanofriction, and in particular of Frenkel-Kontorova-like models. 
This is in particular the case when a topological defect is created in a zigzag ion Coulomb crystal, which results in an Aubry transition from free sliding to pinned phase as a function of the trap aspect ratio. 
We explore the quantum effects of the Aubry transition by means of an effective simplified model, in which the defect is treated like a single quantum particle that experiences an effective Peierls-Nabarro potential and a position-dependent mass. 
We demonstrate the relevance of quantum tunneling in a finite range of aspect ratios close the critical point, showing that the quantum effects may be observed in the kink dynamics for sufficiently low temperatures.  
Finally, we discuss the requirements to reveal quantum effects at the Aubry transition in future experiments on trapped ions.

\end{abstract}

\maketitle


\section{Introduction}

Friction is a relevant phenomenon for a wide range of macroscopic and microscopic systems \cite{VANOSSI2013,GAO2004,KUHNER2007}. 
While its deceleration effect when two macroscopic objects slide on top of each other is phenomenologically described by the laws of Amontons and Coulomb \cite{AMONTONS1699,COULOMB1821}, 
the physics of nanofriction on an atomic scale is much more intriguing \cite{MO2009}. 
One of the simplest and best-known models on nanofriction is the Frenkel-Kontorova (FK) model, in which a chain of particles, which interact through a harmonic potential, is subjected to a sinusoidal corrugation potential \cite{KONTOROVA1938,BRAUN2004}.
The model is hence governed by the interplay between two length scales, the equilibrium distance of the particles and the period of the corrugation potential.  
This interplay leads to an intriguing physics, relevant to disparate fields ranging from solid-state physics \cite{DIENWIEBEL2004,BRAZDA2018,BOHLEIN2012} to biophysics \cite{ENGLANDER1980,BORMUTH2009} and nonlinear physics \cite{CHAIKIN1995,LEPRI2003}. 

Interestingly, if the ratio of the two lengths is incommensurate there exists a transition, the so-called Aubry transition, from a free sliding phase for small potential amplitudes to a pinning phase for strong potentials \cite{AUBRY1983}.
For weak corrugation potentials the classical minimal energy configuration has constant distances given by the inter-particle interaction. 
For incommensurate ratios this has the consequence that particles with arbitrary positions relative to the potential can be found. As a consequence the 
chain of particles may freely slide over the corrugation \cite{SACCO1978}.
In contrast, when the potential amplitude is increased above a critical value the atoms slide towards the minima of the potential getting pinned by the substrate \cite{SHARMA1984,BRAIMAN1990}.
The picture changes when the FK model is treated quantum mechanically, since particles may travel into a neighboring potential well via tunneling effects. 
The behavior of the FK model in the quantum regime has been studied in numerous works, employing Monte Carlo methods \cite{BORGONOVI1989,HU2000,ZHIROV2003} and density matrix renormalization group calculations \cite{MA2014,MA2014a}. 
Whereas the classical pinning phase is characterized by a non-analytic step-wise Hull function due to the presence of Peierls-Nabarro barriers \cite{BAK1982}, quantum tunneling result  
in the softening of the Hull function \cite{BORGONOVI1990a,BERMAN1994,BERMAN1997}. 

Following first proposals \cite{GARCIA-MATA2007,BENASSI2011}, recent experiments have demonstrated the ability to study nanofriction in trapped-ion setups.
Subsequently, two main approaches to study Aubry physics have been realized. The first one builds on a one-dimensional chain of ions confined in a stiff Paul trap \cite{BYLINSKII2015,BYLINSKII2016}. 
They are exposed to a nonlinear substrate potential via the dipole forces of a standing wave laser. 
Nanofriction has then been characterized by the forces necessary to dislocate the particles, revealing the absence of a friction force in the sliding regime. 
A different setup emulates Aubry physics in a self-organized system, that is the two sub-chains of an ion crystal under the presence of a topological defect, which provides the necessary incommensurability between the length scales of the subchains \cite{KIETHE2017a,KIETHE2018}. 
Experiments on this approach have demonstrated the softening of a motional mode at the transition point by measuring its resonance frequency.  
Recent work has explored the relevance of quantum effects in trapped ion experiments on Aubry physics \cite{BONETTI2021}. 
It explicitly revealed the splitting of ion wavefunctions at the Aubry transition when the crystal is cooled close to the ground state and investigated the effect of incommensurability in the system.

In this paper we provide an alternative approach for the study of quantum effects in the FK model.
In our approach, we derive close to the Aubry transition an effective collective description, that permits a considerably simplified analysis of quantum effects. 

This approach is generally applicable to FK systems.
As an example we focus in this paper on a trapped ion
realizations with a topological defect.
We show that quantum effects are important at the Aubry transition, and discuss how the quantum regime may be enlarged in experiments. We discuss as well the case of finite temperature, discussing the 
requirements for the observation of quantum effects. 

This paper is structured as follows. In Sec.~\ref{sec:Aubry} we briefly review the Aubry transition in two dimensional ion crystals. 
In Sec.~\ref{sec:Potential} we derive the effective single-particle model for the motion of a topological defect in the atomic chain. In Sec.~\ref{sec:QuantumKink} we quantize
the kink motion, and analyze the characteristics of the eigenstates of the defect. Section~\ref{sec:Signatures} is devoted to signatures of quantum effects across the Aubry transition and the requirements for their observation.
In Sec.~\ref{sec:Experiments} we discuss the experimental strategies to reveal the quantum effects of the defect. Finally, in Sec.~\ref{sec:Conclusions} 
we summarize our results.


\section{Aubry transition}
\label{sec:Aubry}

In this section, we briefly introduce the Aubry transition in a trapped ion crystal with a topological defect. We consider $N$ ions in a harmonic trap. 
The overall potential experienced by the ions is composed by the trap potential and the Coulomb interaction
\begin{equation}
V(\{\vec r_i = (x_i,z_i)\})=\sum_i^N 
\left(z_i^2+\alpha^2x_i^2\right) 
+\sum_{j\neq i}\frac{1}{|\vec{r}_i-\vec{r}_j|},
\label{eq:Hamiltonian}
\end{equation}
where $(x_i,z_i)$ is the position of the $i$--th ion.  
Hamiltonian~\eqref{eq:Hamiltonian} is written in a dimensionless form, 
employing as energy unit $u=m\omega_zl^2$ and length unit  $l=\left ( C/m\omega_z^2 \right )^{1/3}$, 
where $m$ is the mass of the ions and $C=e^2/4\pi\epsilon_0$ is the Coulomb potential strength with the elementary charge $e$ and the vacuum permittivity $\epsilon_0$. 
The trap is characterized by the secular frequency of the trap in the axial~($z$) direction, $\omega_z$ and by the aspect ratio between transversal and axial frequencies $\alpha = \omega_x/\omega_z>1$. 
In this following we fix {$\omega_z=2\pi\cdot\SI{150}{\kilo\hertz}$}.

For a vanishing temperature the ions crystallize in their equilibrium positions $\vec{r}_i^0$, which in turn depend on $N$ and $\alpha$. For a strong transverse confinement 
the ions are placed in a single chain, i. e. $x_i^0 = 0$, with a non-uniform spacing along $z$ due to the harmonic confinement. For a decreasing $\alpha$ the crystal undergoes a structural linear-to-zigzag phase transition at a critical value $\alpha_\text{ZZ}$. 
Due to the inhomogeneous spacing, the transition occurs first at the trap center in the region of maximal charge density. This configuration is twofold degenerate~(zigzag/zagzig) due to the mirror symmetry $x\leftrightarrow -x$ of the potential.

When $\alpha$ is tuned below a value $\alpha_K$ the zigzag chain can host a kink defect, i.e. a domain wall between a zigzag and a zagzig region, connecting them in a continuous way~(see Fig.~\ref{fig:Crystal}). 
Crucially, due to the presence of the defect, the axial distance in the upper and lower subchains differ. As a result, 
the kink emulates an incommensurate interface of two particle chains, whose influence on one another is determined by the Coulomb repulsion \cite{KIETHE2018}. 
For low values of $\alpha$ the defect is captured in the trap center due to its repulsion from the crystal boundaries \cite{PARTNER2013}. When $\alpha$ is increased the soliton undergoes a Aubry-type transition at $\alpha_\text{A}$, because upper particles are pinned by the periodic Coulomb potential of the lower ones and vice versa.  
The Aubry transition results in spontaneous breaking of the mirror symmetry as the kink equilibrium position shifts away from the trap center \cite{TIMM2020}. 

\section{Effective Potential}
\label{sec:Potential}

\begin{figure}
\includegraphics[width=0.45\textwidth]{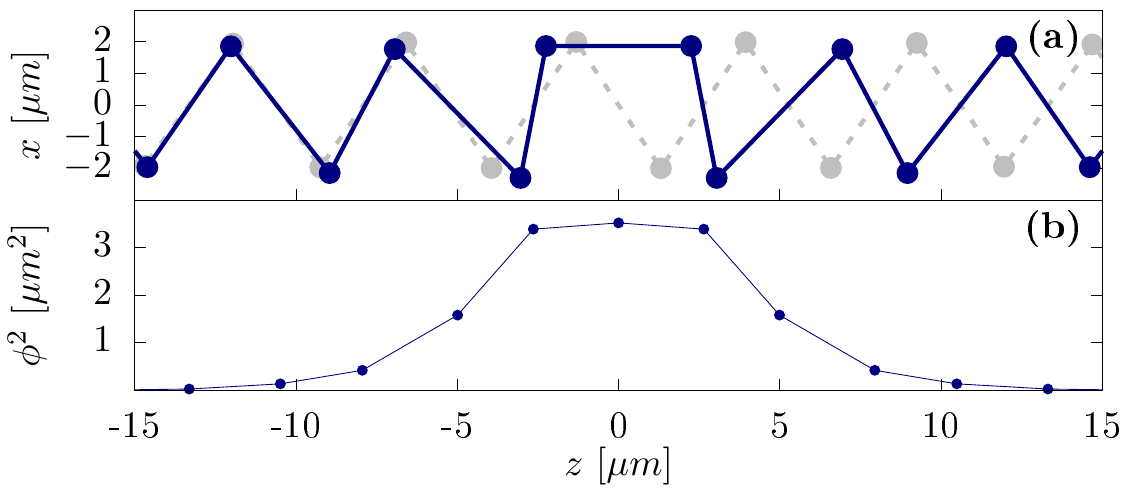}
\caption{\textbf{(a)} Center section of an ion crystal with a topological defect~(blue) and in a defect-free zigzag alignment~(grey) for $\alpha-\alpha_\text{A} = -0.014$. The points are connected to guide the eye. \textbf{(b)} $\phi^2 = (\delta-\delta^{zz})^2$~(see text) for the same choice of parameters.}
\label{fig:Crystal}
\end{figure}

In the following, 
we derive an effective model to describe the kink dynamics. 
The presence of the kink results in the deviation of the axial ion distances from the defect-free case~(see Fig.~\ref{fig:Crystal}). We therefore define the position variable of the kink $K$ as 
\begin{equation}
K (\{\vec r_i\}) = \frac{1}{X_0}\sum_j^{N-1} \bar z_j(\delta_{j}-\delta_j^{zz})^2.
\label{eq:KinkPos}
\end{equation}
Here $\delta_j = z_{j+1}-z_j$ is the $z$-distance between neighboring ions ($\delta_j>0,~\forall j$, as we assume the ions are labelled in $z$-ordering), $\delta_j^{zz}$ is the respective distance with no kink present, $\bar z_j = (z_{j+1}+z_j)/2$, and $X_0=\sum_j(\delta_{j}-\delta_j^{zz})^2$ is a normalization factor. 
The kink position is hence the average axial ion position
weighted with the deviation from the regular zigzag configuration. An example for these deviations is depicted in Fig.~\ref{fig:Crystal} (b). 

The dynamics of the kink position is determined by the Peierls-Nabarro (PN) potential \cite{PEIERLS1940,NABARRO1947}.
To map out the potential landscape we 
minimize the potential energy of the system under the constraint $K(\{\vec r_{i} \}) = X$, for different positions $X$.
The minimization yields ion positions $\vec r_{i,C}(X)$ that determine a path in phase space \cite{PARTNER2013}. The PN potential of the solitonic defect is then defined as the potential energy along that path
\begin{equation}
U_\text{PN}(X) = V(\{\vec r_{i,C}(X)\})
\label{eq:PNPotential}
\end{equation}

\begin{figure}
\includegraphics[width=0.45\textwidth]{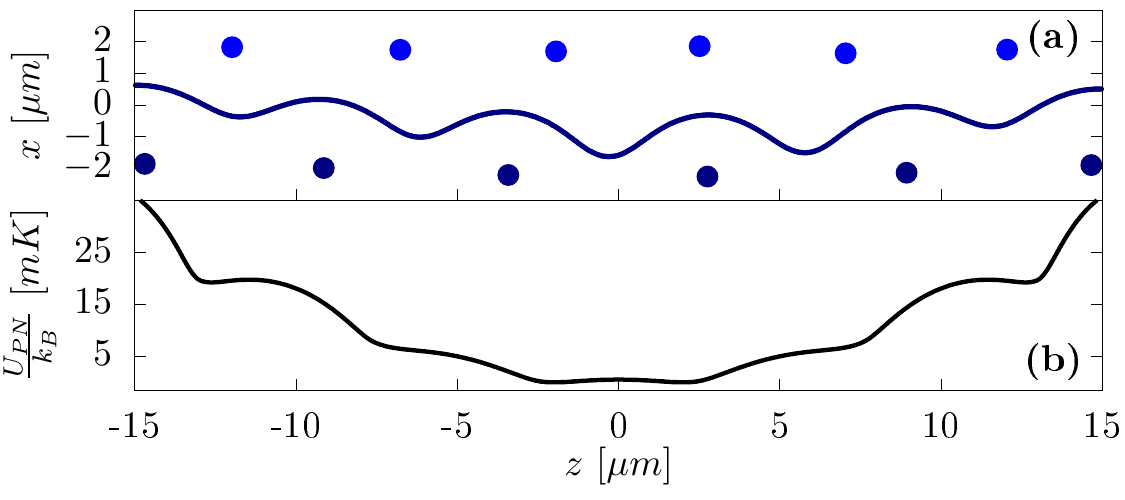}
\caption{\textbf{(a)} Central part of an ion crystal with a kink, composed of the upper (light blue) and lower (dark blue) subchain, and an equipotential line of the corrugation potential of the lower sub-chain felt by the upper ions. 
\textbf{(b)} Effective PN potential of the kink, calculated as derived in section \ref{sec:Potential}, for $\alpha-\alpha_\text{A} = 0.191$. The overall confining potential exhibits periodic modulations due to the influence of the two sub-chains on each other (see (a)). Close to the critical point only the central two local minima are of relevance.}
\label{fig:CorrugationPot}
\end{figure}

\begin{figure}
\includegraphics[width=0.45\textwidth]{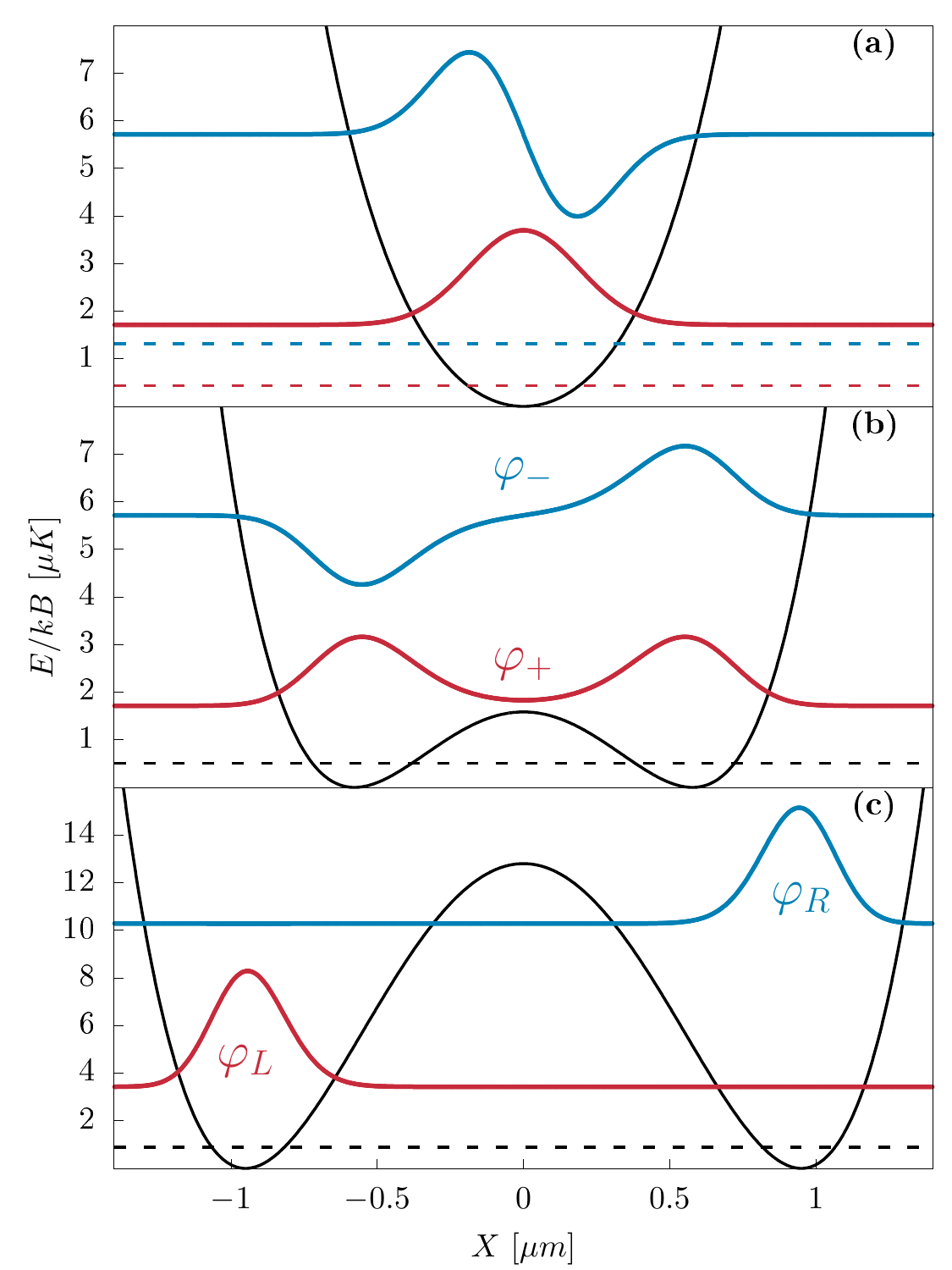}
\caption{PN potential (black) as described in section \ref{sec:Potential} and wavefunctions of the ground state~(red) and the first excited state~(blue), calculated from the model \eqref{eq:KinkH},  for  $\alpha-\alpha_\text{A} = -0.014$ (a), $0.011$ (b) and $0.031$ (c). The graphs of the wave functions have a vertical offset for better visibility. The dashed lines indicate the energy of the shown states, in (b) and (c) the states are quasidegenerate.}
\label{fig:PNPotential}
\end{figure}

The PN potential is shaped by two major effects. On one hand, in the considered parameter regime the defect is repelled by crystal boundaries, where the transversal distance of the ions decreases.
As a result, the potential is globally confining such that the kink remains trapped in the central region of the system. 
On the other hand, the repulsion of the two ion subchains is responsible for a sinusoidal modulation of the potential due to the periodic array of the ions in the crystal, see figure \ref{fig:CorrugationPot}. 
For $\alpha<\alpha_\text{A}$ the transverse ion distances are large such that the repulsion of the subchains is dominated by the repulsion of the boundaries.
Hence, the PN potential of the soliton is practically harmonic with a global minimum at $X_0=0$, which is the classical equilibrium of the kink, see figure \ref{fig:PNPotential} (a).
At $\alpha_\text{A}$ however, the subchains are forced so close together such that the increased Coulomb repulsion causes the PN potential to develop a periodic modulation which leads to several local minima (see Fig.~\ref{fig:CorrugationPot}(b)).
In particular, in the system center the potential resembles a double well, as a local maximum emerges at $X=0$ while two local minima form at $\pm X_0$. 
Classically, the symmetry of the crystal is spontaneously broken as an infinitesimal perturbation leads to a new equilibrium position $X_0 \neq 0 $ of the soliton. 
As we will see in section \ref{sec:QuantumKink} this is not the case for the quantized kink defect.
Increasing $\alpha$ further away from $\alpha_\text{A}$ leads to an increase of $X_0$ as well as a quadratic increase of the potential barrier between the two degenerate local minima $X_0$ and $-X_0$, see Fig.~\ref{fig:PNPotential} (c). 

The effective mass of the kink along the minimization path in phase space is given by
\begin{equation}
M_\text{eff}(X) = m\sum_i \left(\frac{d\vec r_{i,C}(X)}{dX}\right)^2
\label{eq:KinkMass}
\end{equation}
and in general is position dependent. 
Note that the description of the kink dynamics by means of the PN potential and the effective mass only considers configurations which lie on the path in phase space given by $\vec r_{i,C}(X)$, and therefore assumes a good isolation from other degrees of freedom of the system.
A full description of the dynamics of the defect in the environment of the crystal would demand the inclusion of excitations orthogonal to the phase space path, which can be done in the frame of a kink dressing approach \cite{WILLIS1986,BOESCH1988}.
However, molecular dynamics simulations have shown that for low enough temperatures the kink motion couples close to the Aubry transition
only weakly to the additional modes of the crystal due to the low frequency of the kink motion \cite{PARTNER2013,LANDA2010}.
Hence, we assume in the following that the kink dynamics is fully described by the PN potential.

\section{Quantum kink soliton}
\label{sec:QuantumKink}

In this section we compute the classical PN potential $U_\text{PN}(X)$ and its associated $M_\text{eff}(X)$ in the way described in section \ref{sec:Potential} and subsequently quantize the kink defect in the PN potential. We write down the Hamiltonian of the kink as 
\begin{equation}
\hat H_s = \hat P \frac{1}{2M_\text{eff}(\hat X)}\hat P + U(\hat X).
\label{eq:KinkH}
\end{equation}
Note that $M=M_\text{eff}(\hat X)$ does not commute with the momentum operator $\hat P$.
We determine the eigenstate wavefunctions $\varphi_i (X)$ and the eigenenergies of the defect by exactly diagonalizing the Hamiltonian \eqref{eq:KinkH}. We are particularly interested in possible tunneling effects close to the Aubry transition and the energy scale which is necessary to resolve these effects.

Due to the double-well shape of the PN potential, above the critical value of $\alpha$ the ground-state wavefunction develops a double-peak structure in the pinned phase, as seen in Fig.~\ref{fig:PNPotential}(b).
In order to quantify the appearance of a double-peaked wavefunction, and hence of significant quantum tunneling through the PN barrier, we determine the Binder cumulant
\begin{equation}
B = 1-\frac{\braket{(\hat X-\braket{\hat X})^4}}{3\braket{(\hat X-\braket{\hat X})^2}^2},
\label{eq:Binder}
\end{equation}
which as recently shown for the case of an 1D ion chain, can adequately characterize the change of the ground-state wavefunction from a single- to a double-peak form~\cite{BONETTI2021}. 
A Gaussian wavefunction is characterized by $B=0$ whereas a double-peak structure leads to a nonzero value of $B$~(with the chosen normalization, $0\leq B \leq 2/3$).

\begin{figure}
\includegraphics[width=0.45\textwidth]{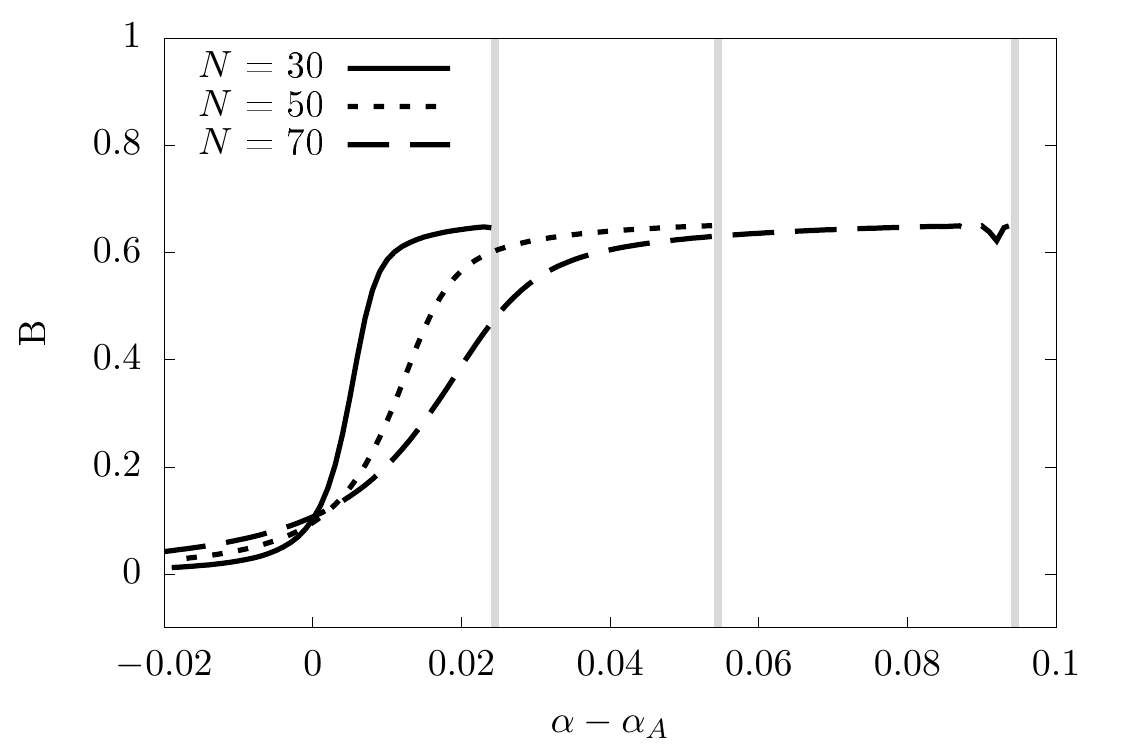}
\caption{Binder cumulant of the ground-state wave function. We show the graph for different particle numbers $N$. The critical point differs for different choices of the particle number: $\alpha_\text{A} (N=30) = 6.409$, $\alpha_\text{A} (N=50) = 9.577$ and $\alpha_\text{A} (N=70) = 12.555$. We only show the graph of the Binder cumulant up to the quasiclasssical regime (see text), marked by the vertical grey bars. }
\label{fig:Binder}
\end{figure}

In Fig.~\ref{fig:Binder} we present the Binder cumulant of the ground state wavefunction as a function of the trap aspect ratio $\alpha$ across the Aubry transition for different $N$. 
For $\alpha<\alpha_\text{A}$ the defect ground state is of Gaussian shape, resulting in $B=0$, as also seen in figure \ref{fig:PNPotential} (a).
At $\alpha=\alpha_\text{A}$ the Binder cumulant increases smoothly to a value of $B=2/3$, indicating non-negligible tunneling effects close to the critical point. 
The low energy states consist of a double peak structure in a symmetric ($\varphi_+$) or antisymmetric ($\varphi_-$) configuration~(see Fig.~\ref{fig:PNPotential} (b)). 
However, as $\alpha$ is increased the distance of the two local minima of the PN potential as well as the energy barrier between them grows. 
As a result, for a sufficiently large $\alpha$ tunneling becomes negligible and the states localized in the left ($\varphi_L$) and the right ($\varphi_R$) potential well decouple from each other, the ground state enters a quasi-classical regime.  
In that regime the Binder cumulant becomes inadequate for reasons discussed in section \ref{sec:Signatures}. 

For a given $N$ the importance of tunneling effects can be characterized by the size of an effective Planck's constant: 
\begin{equation}
\tilde\hbar = \frac{\hbar\omega_z}{m\omega_z^2l^2}=\hbar\left(\frac{\omega_z}{mC^2}\right)^{1/3}
\end{equation}
comparing the quantum energy scale $\hbar\omega_z$ with the classical counterpart $m\omega_z^2l^2$.
Hence, the size of the $\alpha$ window with considerable tunneling effects can be tuned by different choices of $\omega_z$ and the ion species, and also by the particle number $N$~(see Fig.~\ref{fig:Binder}).
Deeper trap potentials or higher number of ions result in shorter ion distances, which increase quantum effects, whereas a smaller ion mass leads to a larger spread of the kink wavefunction, enhancing quantum tunneling.


\section{Signatures of quantum effects}
\label{sec:Signatures}
\begin{figure}
\includegraphics[width=0.45\textwidth]{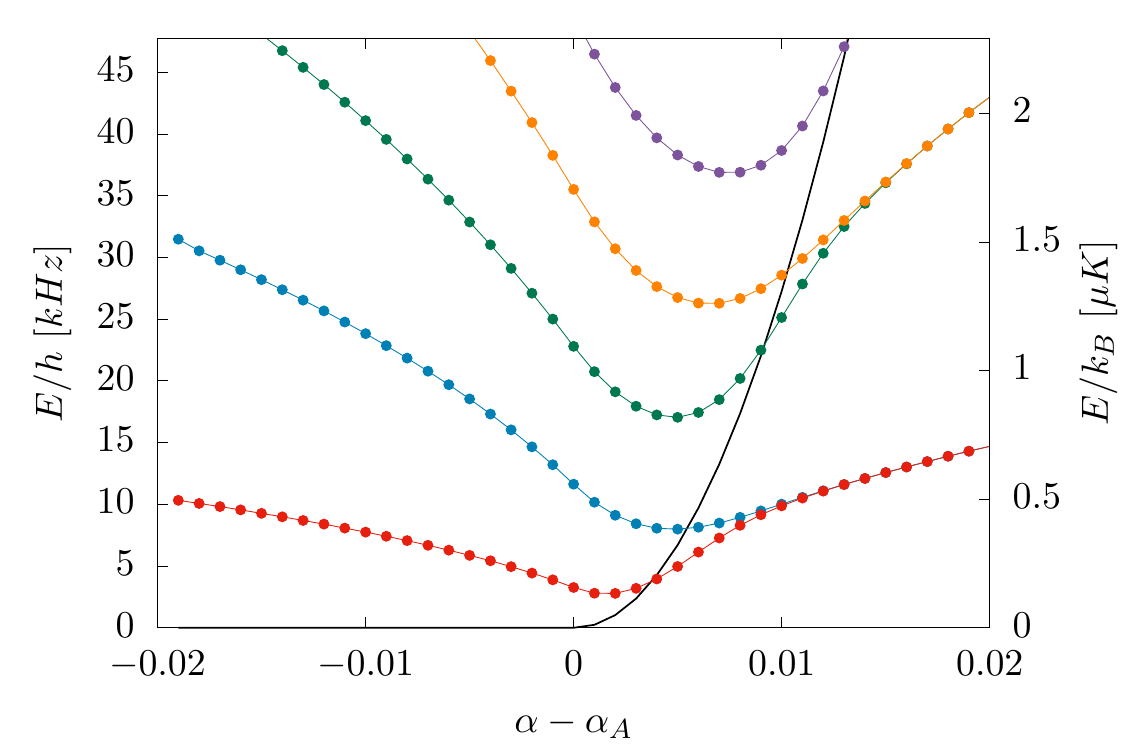}
\caption{Low energy spectrum of the effective kink Hamiltonian across the classical transition point $\alpha_\text{A}$ for $N=30$. The black line indicates the energy barrier of the PN potential $\Delta$. }
\label{fig:Energies}
\end{figure}

As we will discuss, the Binder cumulant alone does not serve as a good measure to assess the presence of quantum effects, therefore we present two alternative ways to reveal the quantum nature of the kink.
Firstly, we present the low-energy spectrum of the quantum kink. 

In the classical regime the transition is marked by a soft mode having vanishing frequency at the critical point \cite{KIETHE2017a}. 
In Fig.~\ref{fig:Energies} we show the low-energy spectrum resulting from exact diagonalization for $N$=30 as a function of $\alpha$. 

Due to quantum fluctuations the energy gap between the ground state and the first excited state remains finite at $\alpha_\text{A}$. 
When $\alpha$ is increased, the eigenenergies of the lowest states become smaller than the PN barrier $\Delta = U(0)-\displaystyle{\min_X U(X)}\propto (\alpha-\alpha_A)^2$~(see Fig.~\ref{fig:Energies}) but maintain their finite gap, indicating the tunneling regime. 

At higher values of $\alpha$ the gap to the first excited state becomes arbitrarily small, resulting in a quasi-degenerate ground-state.
States with higher energies show the same behavior at subsequently larger values of $\alpha$, forming pairwise degenerate states. 
The loss of the energy gaps can be again associated to the vanishing of tunneling effects, leading to spatially separated states located at each potential minimum~(see Fig.~\ref{fig:PNPotential}). 
This is the regime where the spontaneous symmetry breaking of the Aubry transition occurs, i.e., any source of decoherence in the system will lead to the spontaneous selection of a linear combination of the degenerate wave functions that is localized either in the left or the right well. However, even for such a classical state, the Binder cumulant remains nonzero, hence it is not a sufficient condition for the appearance of the quantum kink.
Note that the quasi-classical regime occurs when the eigenenergies of the pairs of states becomes significantly lower than $\Delta$~(see Fig.~\ref{fig:Energies}). 
In contrast, the quantum kink is characterized by the existence of a finite tunneling gap, i.e., at sufficently low temperature, spontaneous symmetry breaking can no longer occur due to the lack of degeneracy. Importantly, the ground state in the quantum kink regime retains the mirror symmetry of the Hamiltonian, which means that is an ordered state (as shown by the finite Binder cumulant) beyond the notion of spontaneous symmetry breaking.

\begin{figure}
\includegraphics[width = 0.45\textwidth]{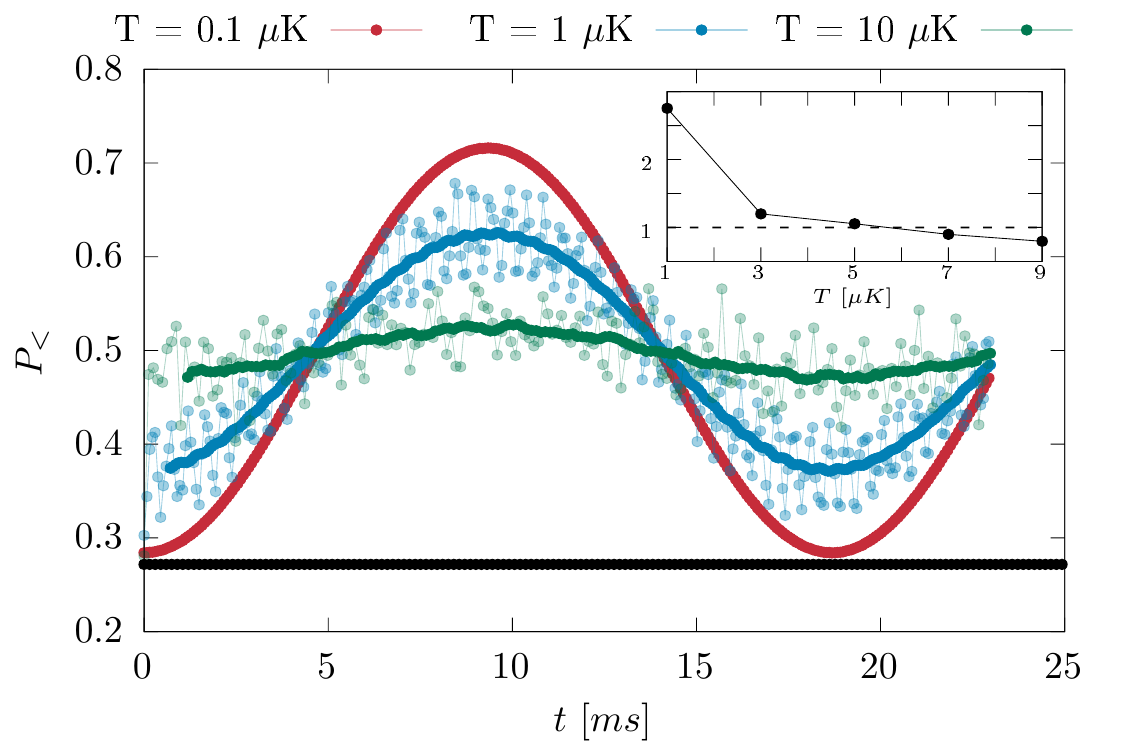}
\caption{Probability to find the kink at $X<0$ as a function of time for $N=30$ and $\alpha-\alpha_\text{A}=0.011$ for different kink temperatures. The kink is prepared in a thermal state of the symmetry broken potential $U(X)-h\cdot X$, where $h/k_BT=0.864~\mu m^{-1}$. The opaque points display a running mean of the dynamics depicted in transparent points. The black line indicates the absence of tunneling for $\alpha-\alpha_\text{A} = 0.031$, even for very small temperatures of $T = \SI{0.1}{\micro\kelvin}$. 
Inset: Ratio between the variance of the running mean and variance of the dynamics within the averaging blocks as a function of the temperature.  }
\label{fig:Experiment}
\end{figure}

Another approach to demonstrate the presence of quantum effects focuses on the different dynamical behavior resulting from tunneling. 

To demonstrate this we initially prepare the kink at one of the potential wells. This is done as follows. We break the mirror symmetry of the PN potential by adding a potential gradient $\hat H_g = -g\hat X$, 
and assume that the system is in a thermal state of the overall Hamiltonian $\hat H = \hat H_s + \hat H_g$.
We choose $g = k_BT/2X_0$ where $X_0$ is the position of the potential minimum and $k_B$ is the Boltzmann constant, such that the potential wells are separated by an energy of $k_B T$.
At $t=0$ we set $g=0$, and monitor the inter-well dynamics. 
Figure~\ref{fig:Experiment} shows the probability $P_<$ to find the kink at $X<0$.
When quantum tunneling is non-negligible $P_<$ oscillates coherently with a frequency given by the energy splitting of the ground state and the first excited state.
When $\alpha$ is tuned into the quasiclassical regime, the kink stays localized in one of the potential wells due to the inability to tunnel through the PN barrier. 
However, for sufficiently large temperatures, the contrast of the oscillations tends eventually to zero, since the kink may move from one well to the other by thermally-activated jumps over the PN barrier, reducing 
the oscillation amplitude and introducing high frequency oscillations in the signal.

To demonstrate this behavior, the trap needs to be stabilized in a configuration which exhibits considerable quantum effects. This demands working with $\alpha/\alpha_A - 1 <  0.002$~(see Fig.~\ref{fig:Binder}) \cite{JOHNSON2016}, although the 
window can be enlarged by increasing the effective Planck's constant~$\tilde\hbar$.
Secondly, the temperature of the defect needs to be well below the energy of the PN barrier in order to preclude thermally activated jumps into the neighboring potential well.
To quantify the necessary temperature scale, we compute running mean values of $P_<$ and comparing the variance of the averaged values with the variance of the data within the averaging blocks (see inset of Fig.~\ref{fig:Experiment}).

The results show that at the chosen $\alpha$ value the kink should be cooled to $T<\SI{5}{\micro\kelvin}$ in order to reveal quantum tunneling of the 
kink. Here again, we want to note that the overall energy scale of the system $E^3=mC^2\omega_z^2$ can be tuned by the trap frequency $\omega_z$, the ion mass $m$ and the particle number $N$, which would influence the necessary temperature scale in turn. 

\section{Experimental accessibility}
\label{sec:Experiments}
In this section we discuss possible strategies to observe the implications of quantum effects in the Aubry transition with trapped ions.
Experimentally reaching the required temperature regime close to the critical point could be achieved by preparing the system deeper in the pinned phase, where the mode frequency of the kink is high enough in order to allow for sufficient cooling while being energetically well isolated from the residual phonon spectrum \cite{LANDA2010}. 
In addition, studies have shown, that heating of short wavelength motional modes by dc electric field noise is strongly suppressed \cite{KALINCEV2021,WINELAND1998,KING1998}, such that effective cooling of the kink dynamics is possible. 
The kink in a crystal of $30$ ions could be prepared at $\alpha-\alpha_\text{A} = 0.091$ where the kink mode exhibits a frequency of $2\pi\cdot\SI{65.17}{\kilo\hertz}$ at an axial trapping frequency of $\omega_z = 2\pi\cdot\SI{150}{\kilo\hertz}$. 
At this frequency, the required temperature of $T < \SI{5}{\micro\kelvin}$ corresponds to a mean motional state occupation of $\bar n < 1.15$. 
After Doppler cooling the entire crystal to the Doppler limit of about $\SI{0.5}{\milli\kelvin}$ cooling can be applied in order to reach a mean occupation number of around $\bar n \lesssim 3 $ for the whole crystal \cite{EJTEMAEE2017,JOSHI2020}, which is close to the desired temperature regime. 
Further cooling of the kink mode can be achieved by incorporating red sideband cooling techniques. 
After the cooling stages, quenching $\alpha$ at a rate of $\dot\alpha\approx-\SI{0.1}{\kilo\hertz}$ ensures adiabaticity during the approach of $\alpha_\text{A}$ while the total quench time on the order of milliseconds is much smaller than the measured lifetime of the defect \cite{PYKA2013}.
Ultimately, the excitation spectrum of the quantum kink presented in Fig. \ref{fig:Energies} can be mapped out in a spectroscopic way.

In an alternative approach aiming to measure the quantum dynamics shown in Fig. \ref{fig:Experiment} the kink needs to be initialized in one of the potential wells. 
This can be accomplished by introducing controllable higher order contributions in the trap potential that break the symmetry in the $z$ direction \cite{BROX2017}.
After the initialization and cooling stage the trap potential can be tuned to be harmonic again in order to trigger the tunneling motion of the kink through the potential barrier. 
This motion translates into an oscillation of the central ions which also takes place on a timescale of tens of milliseconds (see Fig. \ref{fig:Experiment} ). 
The oscillation has an amplitude on the order of a few hundred
nanometers, and requires a non-destructive detection without heating.


\section{Conclusions}
\label{sec:Conclusions}
In summary, we have discussed an effective model for the dynamics of a topological defect in an ion Coulomb crystal. 
Working with an effective potential and effective mass of the kink yields a simplified single-particle description of the relevant physics at the Aubry transition. 
Our results for the quantum kink Hamiltonian demonstrate the presence of quantum effects close to the critical Aubry point, which manifest themselves, 
in the coherent oscillation of the kink between the two minima of the PN potential. 
The observation of quantum tunneling of the defect demands in typical experiments $T<\SI{5}{\micro\kelvin}$.
Sub-Doppler cooling the motion of the kink in ion Coulomb crystals to reach these low temperatures will be a challenge for future experiments on quantum Aubry physics based on trapped ions.
Finally, we would like to mention that 
our effective single-particle model can be directly applied to other systems that exhibit Aubry physics.
In particular, with an appropriate definition of the kink position variable $K({\vec r_i})$, it can be applied to an alternative approach to emulate nanofriction in a trapped ion setup based on a linear ion string in a superimposed periodic lattice, as recently investigated in \cite{BYLINSKII2015,BONETTI2021}. \\

\acknowledgments
We thank Jan Kiethe for fruitful discussions.
This project has been funded by the Deutsche Forschungsgemeinschaft (DFG, German Research Foundation) under Germany’s Excellence Strategy - EXC-2123 QuantumFrontiers - 390837967 and through CRC 1227 (DQ-mat), project A07.
This project 17FUN07 CC4C has received funding from the EMPIR programme co-financed by the Participating States and from the European Union’s Horizon 2020 research and innovation programme. 

\bibliography{QuantumAubryPaper.bib}

\end{document}